\documentclass[12pt,a4paper]{article}
\pdfoutput=1
\usepackage[latin1]{inputenc}
\usepackage{amsfonts,amsbsy,bm,euscript,mathrsfs}
\usepackage{amssymb,stmaryrd,faktor,slashed}
\usepackage{color}
\usepackage[tbtags]{amsmath}
\usepackage[bookmarks=true,colorlinks=true,linkcolor=black,citecolor=black,urlcolor=black,bookmarksnumbered]{hyperref}

\usepackage[a4paper,text={170mm,257mm},centering]{geometry}

\numberwithin{equation}{section}

\makeatletter
\renewcommand\section{\@startsection {section}{1}{\z@}%
{-3.5ex \@plus -1ex \@minus -.2ex}%
{2.3ex \@plus.2ex}%
{\normalfont\large\bfseries}}
\renewcommand\subsection{\@startsection{subsection}{2}{\z@}%
{-3.25ex\@plus -1ex \@minus -.2ex}%
{1.5ex \@plus .2ex}%
{\normalfont\normalsize\bfseries}}
\makeatother

\expandafter\def\expandafter\bfseries\expandafter{\bfseries\ifmmode\else\boldmath\fi}
\expandafter\def\expandafter\mdseries\expandafter{\mdseries\ifmmode\else\unboldmath\fi}
\expandafter\def\expandafter\normalfont\expandafter{\normalfont\ifmmode\else\unboldmath\fi}

\providecommand{\href}[2]{#2}
\newcommand{\arxivlink}[1]{\href{http://arxiv.org/abs/#1}{[arXiv:#1]}}
\newcommand{\doilink}[2]{\href{http://doi.org/#2}{#1}}

\newcommand{\mathsym}[1]{{}}
\def\id{\protect{{1 \kern-.28em{\rm l}}}}
\def\be{\begin{eqnarray}}
\def\ee{\end{eqnarray}}

\def\ha{\tfrac{1}{2}}
\def\td{\tilde}

\def\z{\zeta}

\def\a{\alpha}
\def\b{\beta}

\def\g{\gamma}

\def\k{\kappa}

\def\det{\hbox{det}}

\def\Tr{{\rm Tr}}

\def\l {\lambda}

\def\O{{\mathcal O}}

\def\foot{\footnote}
\newcommand{\rf}[1]{(\ref{#1})}

\def\no{\nonumber}

\def\la{\label}
\def\l{\lambda}

\def\p{\phi}

\def\varpi{{\rm w}}

\def\del{\partial}
\def\s{\sigma}

\def\ed{\end{document}}

\def\iffa{\iffalse}

\def\te{\textstyle}

\def\rk {{\rm u}}

\def\sm{$\sigma$-model }
\def\sms{$\sigma$-models }
\def \ov {\over}
\def\Ad{\text{Ad}}

\def \k {\varkappa} 
\def \dt {\dt}

\def \kk {{\rm k}}
\def \ka {{\kappa}}

\def \cM {\mathcal{M}}

\begin{document}

\begin{flushright}\small{Imperial-TP-AT-2019-{05}}

\end{flushright}

\vspace{2.5cm}

\begin{center}

{\Large\bf Integrable 2d sigma models:\\
\vspace{0.3cm}
quantum corrections to geometry  from  RG flow
}

\vspace{1.5cm}

{Ben Hoare$^{a,}$\footnote{bhoare@ethz.ch}, \  Nat Levine$^{b,}$\footnote{n.levine17@imperial.ac.uk} and
Arkady A. Tseytlin$^{b,}$\footnote{Also at the Institute of Theoretical and Mathematical Physics, MSU and
Lebedev Institute, Moscow.

\ \ tseytlin@imperial.ac.uk}
}

\vspace{0.5cm}

{\em
\vspace{0.15cm}
$^{a}$ETH
Institut f\"ur Theoretische Physik, ETH Z\"urich,\\
\vspace{0.05cm}
Wolfgang-Pauli-Strasse 27, 8093 Z\"urich, Switzerland.
\\
\vspace{0.15cm}
$^{b}$Blackett Laboratory, Imperial College, London SW7 2AZ, U.K.
}
\end{center}

\vspace{0.5cm}

\begin{abstract}
Classically integrable $\sigma$-models are known to be solutions of the 1-loop RG equations, or ``Ricci flow", with only a few couplings running. In some of the simplest examples of integrable deformations we find that in order to preserve this property at 2 (and higher) loops the classical $\sigma$-model should be corrected by quantum counterterms. The pattern is similar to that of effective  $\sigma$-models associated to gauged WZW theories. We consider in detail the examples of the $\eta$-deformation of $S^2$ (``sausage model") and $H^2$, as well as the closely related $\lambda$-deformation of the $SO(1,2)/SO(2)$ coset. We also point out that similar counterterms are required in order for non-abelian duality to commute with RG flow beyond the 1-loop order.
\end{abstract}

\newpage

\tableofcontents

\def \OO {{\cal O}}
\def \etm {$\eta$-model\ }
\def \lam {$\l$-model\ }
\def \dt {\frac{d}{dt}}
\def \rrr {x} \def \qp {{\rm q}}
\def \qp {{\mathrm h} }
\def \kk { k}
\def \bq {{q'}} \def \barp {x} \def \barq {y}
\def \rk {{\rm k}}

\setcounter{footnote}{0}
\setcounter{section}{0}

\section{Introduction}\label{intro}

The bosonic 2d $\s$-model\foot{Here we ignore the $B$-field coupling for simplicity and
absorb the loop-counting parameter
$\hbar=\a'$ into the target space metric $G_{mn}$.}
\be \la{1}
S={1\ov 4\pi } \int d^2 z\,
G_{mn}(x)
\, \del^a x^m \del_a x^n
\ee
is a theory with an {\it infinite} number of couplings
(as can be seen, e.g., by Taylor expanding $G_{mn}(x) = \delta_{mn} + \sum_r g_{mn; k_1 \cdots k_r} x^{k_1} \cdots x^{k_r} $)
that run with RG scale according to the ``generalized Ricci flow" equation \cite{ecker,friedan}
\begin{align}
\dt
G_{mn}
&= \b_{mn} + \nabla_{(m} \xi_{n)}
~, \la{2}\\
\b_{mn} &= R_{mn} + \ha R_{mpqr} {R_n}^{pqr} + \ldots \ .\la{3}
\end{align}
Here $\xi^n(t)$ corresponds RG scale dependent diffeomorphisms, or equivalently field redefinitions of $x^n$.

In certain special cases, the RG flow may be consistently restricted to a {\it finite} subset of couplings.
These \sms may be called {renormalizable} in the usual sense.
In the simplest case of homogeneous spaces, for example group or symmetric spaces, with non-abelian global symmetry,
there is just one running coupling related to the curvature radius.
More non-trivial examples include models related by T-duality, for which the global symmetry becomes partly hidden.
Starting, e.g., with the \sm on $S^2$ (of radius $\sqrt h $)
and T-dualising along the isometry direction $\p$ one finds a dual metric
\be \la{4}
ds^2 = h ( d\theta^2 + \sin^2 {\theta}\, d\p^2 ) \ \ \to \ \
\widetilde {ds^2} = h ( d\theta^2 + \frac{1}{\sin^2 {\theta} }\, d\tilde{\p}^2)\ee
that no longer has a global $SO(3)$ symmetry. Instead, this symmetry
is hidden as a consequence of integrability
\cite{ricci} and the dual metric still solves the 1-loop RG equations with only one coupling $h(t)$.

More generally, it is now widely believed that
renormalizability, or invariance under the RG flow,
is closely linked with the integrability of a \sm \cite{Fateev:1992tk,fateev96,lukyanov}
(for some recent examples see, e.g., \cite{Fateev:2019xuq,Fateev:2018yos,Demulder:2017zhz,Klimcik:2019kkf}).
One motivation for this, implicit in \cite{Fateev:1992tk,fateev96}, is that a \sm corresponding to a given a quantum integrable S-matrix with a finite number of parameters should also be parametrized by only a finite number of couplings. Another is that the conservation of infinitely many hidden symmetry charges
should be enough to reduce the infinite-dimensional RG flow to a finite-dimensional subspace.

While there are no known counterexamples
to the conjecture that classically integrable \sms should be invariant under RG flow, in most examples
this has only been checked at the leading 1-loop
order
in \rf{2},\rf{3}, and there is no general proof.

\medskip

The aim of the present paper is to address what happens beyond the leading
1-loop order in some of the simplest non-trivial examples of integrable \sms with 2d
target spaces without non-abelian global symmetry.
We shall see that the classical \sm metric should generally be deformed in order to remain a solution of the higher-loop RG equation \rf{1} with the same number of running couplings.
This order-by-order deformation of the metric may be interpreted as a result of adding local ``counterterms" required to preserve the integrability at the quantum level (i.e. to satisfy ``Ward identities"
for hidden symmetries).
Indeed, in cases with a UV fixed point and a known underlying quantum S-matrix,
the deformed \sm may be reconstructable from a dual massive model (cf. \cite{fateev96,Litvinov:2018bou}).

One example where such a deformation is expected is the model defined by $\widetilde {ds^2}$ in \rf{4}, which is T-dual to the \sm on $S^2$. Given the metric
$ds^2 = h [dx^i dx^i + \cM(x) dy^2] $ the standard T-duality rule $\td \cM= \cM^{-1}$ is known to be
modified at the 2-loop level \cite{Tseytlin:1991wr} (see also \cite{Panvel:1992he,Haagensen:1997er,Haagensen:1997bh})
\be
\td \cM = \cM^{-1} \big( 1 + \tfrac{1}{2} h^{-1} \, \del_i \log \cM\, \del^i \log \cM \big) \ , \la{5}\ee
from which one can straightforwardly determine the corresponding quantum correction
to the dual metric in \rf{4}.\foot{In \rf{4} we have
$x=\theta$ and $\cM= \sin^2 \theta$ so that \rf{5} implies
$\csc^2 \theta \to \csc^2\theta( 1+ 2 h^{-1} \cot^2 \theta + \OO (h^{-2}))$.
}

Another known example of such quantum corrections is provided
by the special integrable \sms corresponding to gauged WZW models.
These models are
scale invariant, i.e. fixed-point solutions of the RG equation \rf{2}.
The conformal invariance of the underlying quantum gWZW theory allows one to find an exact
form of the \sm metric (and $B$-field) \cite{Dijkgraaf:1991ba,Bars:1992sr,Tseytlin:1992ri,bs,Tseytlin:1993my}.\foot{This is true in a particular scheme in which the ``tachyon" equation is not modified
(for a discussion and examples see \cite{Tseytlin:1993df,Horowitz:1994ei,Tseytlin:1995fh}).
Here the level $k$ plays the role of the inverse coupling (and loop-counting parameter) $h$ used above.}
The simplest example is provided by the exact counterpart \cite{Dijkgraaf:1991ba} of the
familiar classical $SL(2,R)/U(1)$ gWZW metric \cite{Bardacki:1990wj,Witten:1991yr}
\begin{align}
ds^2 = k(dr^2 + \tanh^2 r \, dy^2) \ \ \to \ \
ds^2 = (k-2) \Big[ dr^2 + \big( \coth^2 r - \tfrac{2 }{k}\big)^{-1} dy^2\Big] \la{6} ~.
\end{align}
As was checked directly in \cite{Tseytlin:1991ht}, the leading $k^{-1}$ correction in \rf{6}
is precisely the one
required to solve the 2-loop scale invariance equation implied by \rf{2},\rf{3}
(see also \cite{Jack:1992mk} for a 4-loop test of \rf{6}).\foot{In this Weyl invariant case
the vector $\xi_n$ can be written as $ 2 \del_n \Phi$ where the exact dilaton field $\Phi$ is given by
$e^{-2 \Phi}= \sinh 2 r \ \big( \coth^2 r - \tfrac{2 }{k}\big)^{1/2}$. \la{f4}}

The quantum deformation of the effective \sm metric associated with gWZW models is
also essential for quantum integrability of generalized sine-Gordon models \cite{FernandezPousa:1996hi}. In particular, as was shown in \cite{Hoare:2010fb} the finite counterterm
required to be added to the action of the complex sine-Gordon model
$L= k \big[(\del x)^2 + \tan^2 x \, (\del y)^2 - m^2 \sin^2 x\big]$
to ensure factorization of the corresponding S-matrix at the 1-loop level
\cite{deVega:1982sh,Bonneau:1984pj} is
precisely the same as the $k^{-1}$ term in the $SU(2)/U(1)$ analog of the exact metric in \rf{6}.

One way to understand the origin of the quantum correction in \rf{6} is
to note that integrating out the 2d gauge field $A_a$ in the gWZW model
leads to a non-trivial determinant that can
be computed exactly \cite{Tseytlin:1991wr,Schwarz:1992te} (see also \cite{Gerasimov:1990fi})\foot{Here the integral is understood to be over 2d scalars $u,v$ defined by $A_+ =\del_+ u, \ A_-=\del_- v$
and having measures $M=M(x(z))$. For generality we assumed a curved 2d background with curvature $R^{(2)}$ (ignoring trivial $M$-independent factor in \rf{7}).
Note that in the discussion of the abelian T-duality transformation at the path integral level in \cite{Buscher:1987qj}
the $\del^a \log M\, \del_a \log M $ term was missed but, in fact, it cancels against similar contribution
coming from another determinant involved \cite{Schwarz:1992te} so that the end result at the leading order is just the dilaton shift found in
\cite{Buscher:1987qj}. The analog of the $\del^a \log M\, \del_a \log M $ term survives, however, in the case of non-abelian duality (see below).}
\be \la{7}
\int [dA] \, e^{ i \int d^2 z\, \sqrt g
\, M\, A^a A_a} = \exp \Big[\frac{i}{4 \pi} \int d^2 x \sqrt g \big(
- \ha \del^a \log M\, \del_a \log M - \ha R^{(2)} \log M \big) \Big]\ .
\ee
This ubiquitous determinant appears also in the discussion of the abelian (and non-abelian) T-duality transformations
with the $R^{(2)} \log M$ term representing the dilaton shift found in \cite{Buscher:1987qj} (see also \cite{DeJaegher:1998pp}).
The local term $ \del^a \log M\, \del_a \log M $ gives a non-trivial 1-loop correction to the classical \sm action
(cf. also \cite{Horowitz:1994ei,Hoare:2010fb}).
This observation will be useful below.

\medskip

In what follows we shall determine the quantum corrections
required to solve the 2- and higher-loop RG equation \rf{2},\rf{3} for the
the simplest
non-trivial integrable models:
the $\eta$-deformation \cite{Klimcik:2002zj,Klimcik:2008eq,Delduc:2013fga}
and the $\l$-deformation \cite{Sfetsos:2013wia,Hollowood:2014rla} of the $S^2$ $\s$-model and its analytic continuations.
In section 2 we will consider the $\eta$-model and, using certain special limits and an analogy with the gWZW model,
will conjecture an exact counterpart of the classical metric. We will confirm that it solves the RG equation \rf{2},\rf{3} to 3 loops.
In section 3 we will consider the $\lambda$-model. Starting from its form as a deformation of the $G/G$ gWZW model, we will
determine the 1-loop correction that comes from integrating out the 2d gauge field $A_a$ and confirm that this solves the 2-loop RG equation \rf{2},\rf{3}.
Since the non-abelian dual of $S^2$ is a limit of the $\lambda$-model, our result implies that accounting for the determinant contribution in \rf{7}
should also resolve past problems \cite{n1,n2,n3,Bonneau:2001za}
in verifying that the non-abelian duality commutes with RG flow at the 2-loop level.

\section{\texorpdfstring{$\eta$}{eta}-model}

The metric of the 2d ``sausage" model
\be
ds^2 = h \Big[ \frac{dr^2}{(1-r^2)(1+\k^2 r^2)} + \frac{1-r^2}{1+\k^2 r^2} d\p^2 \Big] \ , \la{21}
\ee
was originally written down in \cite{Fateev:1992tk} as the leading semi-classical approximation to the \sm corresponding to the massive integrable trigonometric S-matrix of \cite{Zamolodchikov:1980ku}. It was discovered as a solution to the 1-loop RG equation \rf{2},\rf{3} and conjectured to be classically integrable. Its classical integrability was later shown in \cite{lukyanov}. In \cite{Hoare:2014pna} this model was identified as the Yang-Baxter \sm \cite{Klimcik:2002zj,Delduc:2013fga}
(also known as the $\eta$-deformation) corresponding to $S^2$.\foot{It is also interesting to note that the metric \rf{21} is formally self T-dual, i.e. invariant under
$\phi\to \td \phi $ and $\td \phi \to i \k^{-1} \phi$,\ \ $r \to i \k^{-1} r^{-1}$.}

Here $\k$ is the deformation parameter, with $\k = 0$ corresponding to the round $S^2$ of radius $\sqrt h$.
The metric \rf{21} has two regimes of interest: real $\varkappa$, which gives the Yang-Baxter \sm\unskip, and imaginary $\varkappa$.
It is the latter case that was studied in \cite{Fateev:1992tk} since it is UV stable with a UV fixed point at $\k^2 = -1$, where the theory is free.
Alternatively, we can send $\k^2\to -1$ while simultaneously expanding around $r^2 = 1$.
Taking the limit in this way the metric \rf{21} reduces to the classical metric of the $\frac{SO(1,2)}{SO(2)}$ gWZW model (cf. \rf{6})
\be r^2 = 1 - (1+\k^2) \sinh^2 \rrr \ , \ \ \qquad \k^2 \to -1 \la{22}\ , \qquad
ds^2\to h( d\rrr^2 + \tanh^2 \rrr\, d\p^2) \ . \ee
Another useful limit is the maximal deformation limit $\k \to \infty$, $h\to \infty$ with $h' \equiv \tfrac{h}{\k^2}$ fixed, which yields 
\cite{Hoare:2014pna} the undeformed hyperbolic space $H^2$  with radius $\sqrt{h'}$ (here we set $r=\tfrac{1}{\cosh{ \theta}}$)
\be
ds^2 = \frac{h'}{r^2} \Big[ \frac{dr^2}{1-r^2} + (1-r^2) d\p^2 \Big] = h' \Big[ d \theta^2 + \sinh^2 \theta \, d\p^2 \Big] \ . \la{222}
\ee

The metric \rf{21} solves the 1-loop RG equation \rf{2},\rf{3} with $h$ and $\k$ running as 
\be \la{23}
\dt h = (1+\k^2) + \OO(h^{-1}) \ , \qquad \qquad
\dt \k = h^{-1} \k (1+\k^2) + \OO(h^{-2}) \ . \ee
We also find that the 1-loop RG flow \rf{23} is effectively 1-coupling as there exists an RG-invariant combination of couplings
\be
\dt \nu =0\ , \qquad \qquad \la{244}
\nu \equiv {h \ov \k} + \O(h^0) \ . \ee
In agreement with the above comments,
$\k^2=-1$ is a UV fixed point.
In the gWZW limit $h$ is identified with level $k$ and hence should not run.

This relation to the gWZW model suggests that,
for \rf{21} to remain a solution to the RG equation at higher loops with only $h$ and $\k$ running,
this metric should be modified by quantum (i.e. $1/h$) terms.
Inspired by the analogy with the exact gWZW metric in \rf{6} we propose
the following conjecture for the exact generalization of \rf{21}
\begin{align}
ds^2 = (h- 1+\k^2 ) \Big[ \frac{dr^2}{(1-r^2)(1+\k^2r^2)} + \big(
\frac{1+ \k^2r^2}{1- r^2} + \frac{2\k^2}{h} \big)^{-1}
d\p^2 \Big] \ . \la{24}
\end{align}
The metric \rf{24} is expected to solve the RG equations to all loop orders
(in a particular scheme and modulo coordinate redefinitions) with $h$ and $\k$ running
according to a generalization of \rf{23}.
This conjecture passes some obvious tests:
(i) the metric \rf{24} reduces to $S^2$ for $\k=0$, now with
shifted radius\unskip\foot{The radius squared parameter ${h-1}$ is related to ${h}$ simply by a finite coupling redefinition.
Alternatively, one may shift $h\to h+1-\k^2$ in \rf{24} to obtain
$ds^2 = h \Big[ \frac{dr^2}{(1-r^2)(1+\k^2r^2)} + \big(
\frac{1+ \k^2r^2}{1- r^2} + \frac{2\k^2}{h+1-\k^2} \big)^{-1}
d\p^2 \Big] ,$
so that the radius of $S^2$ is $h$ in the $\k=0$ limit.} $\sqrt{h-1}$; (ii) the metric remains flat for $\k^2=-1$;
(iii) in the non-trivial $\k^2 \to -1$ limit \rf{22} it reduces to the exact gWZW metric in \rf{6} with level $k=h$; (iv) in the maximal deformation limit \rf{222} it reduces to $H^2$, now with shifted radius $\sqrt{h'+1}$.

More importantly,
one can directly check that \rf{24} with the leading $h^{-1}$ correction
included
\begin{align} \ \no ds^2 &= h \Big[ \frac{dr^2}{(1-r^2)(1+\k^2r^2)} \big(1-{{1-\k^2}\ov{h}}\big) \\
&\qquad \qquad\qquad \quad + \frac{1- r^2}{1+ \k^2 r^2} \big(1 - {{1-\k^2}\ov{h}} -{ 2 \k^2 \ov {h} } \frac{1- r^2}{1+ \k^2 r^2} + \ldots \big) d\p^2 \Big] \ , \la{277}
\end{align}
indeed solves the 2-loop RG equation\foot{Note that in
2 dimensions $R_{mnkl} = \ha R ( G_{mk} G_{nl} - G_{nk} G_{ml})$.}
\rf{2},\rf{3}
\be \la{27}
\dt G_{mn} = (\ha R + \tfrac{1}{4} R^2 ) G_{mn} + \ldots + \nabla_{(m} \xi_{n)} \ ,
\ee
for a particular diffeomorphism vector $\xi_n$ with components\foot{We note that the diffeomorphism vector \rf{2777} begins at 2-loop order; the 1-loop diffeomorphism vanishes in these coordinates.}
\be \la{2777}
\xi_\p = 0 \ , \qquad \quad \xi_r = \frac{2\k^2 (1+\k^2) r}{h (1+\k^2 r^2)^2} \ .
\ee
The $\beta$-functions \rf{23} receive the following 2-loop corrections
\begin{align}
\la{28}
\dt h = (1+\k^2)&\big[1 + h^{-1} (1-\k^2) + \OO(h^{-2})\big] \ , \\
\la{288} \dt \k = h^{-1} \k (1+\k^2)&\big[1 + h^{-1} (1-\k^2) + \OO(h^{-2}) \big] \ .
\end{align}
The RG-invariant quantity in \rf{244} remains RG-invariant without 2-loop corrections
\be
\dt \nu =0\ , \qquad \qquad
\nu \equiv {h \ov \k} + 0 + \mathcal{O}(h^{-1}) \ . \la{2888}\ee
Furthermore, the ansatz \rf{24} also solves the 3-loop RG equation \rf{2} with
\be \la{29}
\b_{mn} = \Big[ \ha R + \tfrac{1}{4} R^2 + c_1 R^3 + c_2 (\nabla R)^2 + c_3 R \nabla^2 R \Big] G_{mn}
+ c_4 \nabla_m R \nabla_n R \ , \ee
in a particular ``natural" renormalization scheme corresponding to
\be \qquad c_1 = 0 \ , \qquad c_2 = \tfrac{1}{8} \ , \qquad c_3 = -\tfrac{1}{4} \ , \qquad c_4=- \tfrac{1}{16}\ ,
\la{31}
\ee
in which the 3-loop $\b$-functions for $h$ and $\k$ take the form
\begin{align}
\la{32}
\dt { h} =(1+\k^2)&\Big[1 +h^{-1} (1-\k^2) + h^{-2} (1-\k^2)^2 + \O(h^{-3}) \Big] \ , \\
\la{33} \dt { \k} = h^{-1} \k (1+\k^2)&\Big[ 1 +h^{-1} (1-\k^2) + h^{-2} (1-\k^2)^2 +\O(h^{-3})\Big] \ .
\end{align}
The diffeomorphism vector in \rf{2777} receives the following 3-loop correction
\be \la{300}
\xi_\p = 0 \ , \qquad \quad \xi_r = \frac{2\k^2 (1+\k^2) r}{h (1+\k^2 r^2)^2} \Big[ 1 + \frac{2\k^2 (r^2-1)}{h(1+\k^2 r^2)} \Big] \ .
\ee
This scheme \rf{31} is related to the minimal subtraction scheme in \cite{Graham:1987ep,Foakes:1987ij,Foakes:1987gg}
\be \la{30}
\qquad c_1 = \tfrac{5}{32} \ , \qquad c_2 = \tfrac{1}{16} \ , \qquad c_3 = 0\ , \qquad c_4=-\tfrac{1}{16}\ ,
\ee
by the covariant coupling redefinition
\begin{align}
G_{mn}^{(\rm nat)}= \big[ G_{mn} + \tfrac{5}{8} \, \a'^2 ({R}^2)_{mn} + \ha \, \a'^2 \nabla^2 R_{mn} \big] ^{(\rm min)}\ ,
\end{align}
where $G_{mn}^{(\rm nat)} \equiv G_{mn}$ is the metric \rf{24} in the ``natural" scheme and $G_{mn}^{(\rm min)}$ is the corresponding metric in the minimal scheme (see Appendix \ref{A}).

We call the scheme \rf{31} ``natural" because the RG-invariant quantity in \rf{2888} remains RG-invariant with no 3-loop corrections:
\be
\dt \nu =0\ , \qquad \qquad
\nu \equiv {h \ov \k} + 0 + 0 + \mathcal{O}(h^{-2}) \ . \ee
This prompts us to conjecture that, in a natural choice of
subtraction scheme at each loop order, the RG-invariant quantity $\nu\equiv \tfrac{h}{\k}$ is an exact RG-invariant.
This suggests that $\nu$ should be the parameter that appears in the exact quantum trigonometric S-matrix
that generalizes the non-perturbative massive S-matrix of the $O(3)$ invariant $S^2$ \sm
\cite{Zamolodchikov:1978xm}.

Relatedly, we extrapolate from the obvious pattern in \rf{32},\rf{33} to conjecture that, in the same natural scheme, the all-loop $\b$-functions of $h$ and $\k$ are
\begin{align}
\dt { h} &=\frac{1+\k^2}{h-(1-\k^2) } \, h \ , \la{exactb1} \\
\dt { \k} &=\frac{1+\k^2}{h- (1-\k^2)} \, \k \ . \la{exactb2}
\end{align}
Here $\k^2=-1$
remains a fixed point, as it should to all orders since it corresponds both to flat space and the gWZW limit.
For $\k=0$ we get $\dt { R^2_{S^2}} =1+R^{-2}_{S^2}$ for the $S^2$ radius $R_{S^2}=\sqrt{h-1}$, which agrees with the 1-loop and 2-loop coefficients in the $\b$-function of the
$S^2$ model. In the maximal deformation limit \rf{222} we get $\dt { R^2_{H^2}} =-1+R^{-2}_{H^2}$ for the $H^2$ radius $R_{H^2} = \sqrt{h'+1} = \sqrt{h\k^{-2}+1}$. The fact that this is correctly related to the $S^2$ $\b$-function by the analytic continuation $R^2_{H^2} = - R^2_{S^2}$ is a
 check of the consistency of the conjectured exact metric \rf{24} and exact $\b$-functions \rf{exactb1},\rf{exactb2}. Moreover, 
 if there is a natural scheme where \rf{exactb1},\rf{exactb2} are exact, 
  then the $S^2$ and $H^2$ $\b$-functions are 2-loop exact in this scheme
   (this may be possible in a special non-minimal scheme  since the 3- and higher loop  $\b$-function coefficients are scheme-dependent).

Let us mention that
in the case of the
(1,1) supersymmetric generalization of the \sm \rf{1}, the first potential correction to the 1-loop $\b$-function
appears
at 4 loops ($\sim \z(3) R^4$ in the minimal scheme \cite{Grisaru:1986px}, with this particular invariant actually vanishing in the case of 2d target space).
There is also
no deformation of the super gWZW metric (to all orders) and it is thus natural to expect that the same will apply to the
model \rf{21} (i.e. the form of the metric will be the same while $h$ and $\k$ will run).

One may also repeat the above discussion for the
$\eta$-deformation of hyperbolic space $H^2= \frac{SO(1,2)}{SO(2)}$
(or the Lorentzian signature symmetric spaces $dS_2$ and $AdS_2$).
The corresponding metric is related to \rf{21} by the
formal analytic continuation
\unskip\foot{This analytic continuation gives the $H^2$ Yang-Baxter \sm based on the split R-matrix of $Lie(SO(1,2))$.
On the other hand, if we had not continued $\k$ we would have found the $H^2$ Yang-Baxter \sm based on the non-split R-matrix.
For $S^2$ only the latter case exists as a Yang-Baxter deformation in a strict sense since there is no split R-matrix of $Lie(SO(3))$ (see, e.g., \cite{BDCGR}).}
\be \la{344}
r\to i r\ , \qquad \ \ \p\to i \p\ , \ \qquad \ \k \to i \k\ , \qquad \ \ h\to - h \ ,\ee
i.e. we find\foot{\la{ftd}Let us note that the metric \rf{34} is self T-dual, i.e. invariant under
$\phi\to \td \phi $ and $\td \phi \to \k^{-1} \phi$,\ \ $r \to \k^{-1} r^{-1}$.}
\be\la{34}
ds^2 = h \Big[ \frac{dr^2}{(1+r^2)(1+\k^2r^2)} + \frac{1+r^2}{1+\k^2 r^2} d\p^2 \Big] \ . \ee
For
$\k=0$ this is the $H^2$ metric, while for $\k^2=1$ it is flat.
The limit analogous to \rf{22}, i.e.
$\k \to 1$ with $ r^2\to -1 + (1-\k^2) \sinh^2 t$, is now a formal limit
and gives the metric
$ ds^2 = h( - dt^2 + \tanh^2 t\, d\p^2)$, which is the classical metric of the
$\frac{SO(1,2)}{SO(1,1)}$ gWZW model.

In
conformally-flat coordinates the metric \rf{34} may be written as
\be \la{35}
ds^2 = {h \ov p^2 + \k^2 q^2} ( dp^2 + dq^2) \ , \qquad \qquad
r = \frac{q}{p} ~, \quad \phi = \ha \log({p^2+q^2})\ ,
\ee
where the scaling symmetry $(p, q) \to \l (p, q)$ is the counterpart of
$\p$-shift isometry of \rf{34}.

The analog of the exact metric \rf{24} is found by the analytic continuation \rf{344}\footnote{It
is interesting to note that the exact metric may have different properties compared to the classical metric.
For example, in the case of the non-split $\eta$-deformation of AdS$_2$   with 
$\k \to  i \k, \     r \to  \rho, \    \phi \to  i t $  the singularity is at 
$\rho^2 = {h - 2 \k^2 \over  \k^2 (h+2)}$, i.e. is always  present in the classical limit $h \gg 1$ 
but is  absent  for $h <  2 \k^2$.}
\be \la{36}
ds^2 = (h+1+\k^2) \Big[ \frac{dr^2}{(1+r^2)(1+\k^2r^2)} + \big(
\frac{1+ \k^2r^2}{1+ r^2} + \frac{2\k^2}{h} \big)^{-1}
d\p^2 \Big]\ . \ee
Again, changing coordinates $ r^2 \to -1 + (1-\k^2) \sinh^2 t$, the limit
$\k^2 \to 1$ is the exact metric of the $\frac{SO(1,2)}{SO(1,1)}$ gWZW model.
Written in the conformally-flat coordinates as in \rf{35}, the quantum-corrected metric is
\be \la{37}
ds^2 = (h+ 1+\k^2 ) \Big[ { dp^2 + dq^2 \ov p^2 + \k^2 q^2} - \frac{\k^2}{2 h} \, \frac{ \big[ d (p^2+q^2)\big]^2 }{(p^2+\k^2 q^2)^2( 1 + {2\k^2 \ov h} { p^2+ q^2\ov p^2 + \k^2 q^2})} \Big]\ .
\ee
Expanding the metric \rf{37} to first subleading order in small $h^{-1} $ one obtains
\be \la{38}
ds^2 = h\, { dp^2 + dq^2 \ov p^2 + \k^2 q^2} + (1+\k^2){ dp^2 + dq^2 \ov p^2 + \k^2 q^2} - \frac{\k^2}{2} \, \frac{ \big[ d (p^2+q^2)\big]^2 }{
( p^2 + \k^2 q^2)^2 } + \OO(h^{-1})
\ . \ee

\section{\texorpdfstring{$\l$}{lambda}-model }

Let us now discuss the quantum deformation of the \sm corresponding to the $\l$-model.
The $\l$-deformation \cite{Sfetsos:2013wia,Hollowood:2014rla} of the $G/H$ symmetric space \sm can be
constructed by starting with the $G/G$ gauged WZW action and adding
a deformation term quadratic in the gauge field that breaks the gauge symmetry to $H$ and has coefficient $b^{-2}$.
Dropping the WZ term for $g\in G$ since it is a total derivative in our case of interest with a 2-dimensional target space,
the Lagrangian is given by
\be \la{40}
L = \kk\, \Tr\big[ \tfrac12 (g^{-1} \del g)^2 + A_+ \partial_- g g^{-1} - A_- g^{-1} \partial_+ g
- g^{-1} A_+ g A_- + A_+A_-
+ b^{-2} A_+ P A_- \big] \ .
\ee
Here $P=P_{G/H}$ is the projector onto the grade 1 part of the algebra $Lie(G)$.

This model has two important limits.
For $b\to0$ the gauge field satisfies the constraint $PA_\pm = 0$ and thus we find the $G/H$ gWZW model.
Setting $g= e^{ v/\kk}$ and taking $\kk\to \infty$, $b\to \infty$ (with $h\equiv \ha \kk\,b^{-2}$ fixed)
gives the model
\be \la{41}
L= \Tr ( v F_{+-} + 2 h A _+ P A_- )\ , \quad \qquad F_{+-} \equiv \del_+ A_- - \del_- A_+ + [A_+, A_-] \ , \quad h\equiv \ha \kk\,b^{-2} \ , \ee
which interpolates between the $G/H$ symmetric space \sm (found by integrating out $v$ and solving $F_{+-} = 0$ by $A_a= f^{-1} \del_a f$) and its non-abelian dual or NAD (found by integrating out $A_a$ to give a \sm for $v$).

Let us consider the simple case of $\frac{G}{H} = \frac{SO(1,2)}{SO(2)}$, for which the $\l$-model is a deformation of the NAD of the \sm on $H^2$.
We fix the $SO(2)$ gauge symmetry by choosing the following parametrization of the coset element
\be \la{42}
\te g = \exp(\alpha \s_3) \exp({i} \beta \s_2) \ , \ \ \ \ \ \ \cosh \alpha = \sqrt{p^2+q^2}\ ,\ \ \ \ \ \tan \beta = \frac{p}{q} \ , \ee
and then solve for the gauge field $A_a$ in \rf{40}.
The resulting \sm is
\be \la{43}
{L} = \frac{\kk}{p^2+q^2-1}\big(\ka\, \partial_+p \partial_-p + \ka^{-1}\partial_+ q\partial_-q\big)\ , \qquad \qquad
\ka \equiv { (1+2b^2)^{-1}}
\ . \ee
The 2d target space metric may be written in a manifestly conformally-flat form by further redefining $q \to \ka q $.

The model \rf{43} again admits several useful limits.
Rescaling the coordinates $ p \to \g p, \ \ q\to \g \kappa q $ and then taking $\g \to \infty$,
the metric becomes equivalent to the \etm metric in \rf{35} \cite{Hoare:2015gda}
with $\ka=\k$.\foot{In general,
the ``boost" limit in the Cartan directions of the \lam gives the T-dual of the \etm \cite{Hoare:2015gda}.
However, the $\eta$-deformation of $H^2$ is self T-dual (cf. footnote \ref{ftd}).}
The case of $\ka=1$ ($b=0$) corresponds to the classical metric of the $SO(1,2)\ov SO(2)$ gWZW model:
\unskip\foot{This metric \rf{45} is formally related to the classical metric in \rf{6} by the imaginary shift $\a \to \a + \tfrac{i\pi}{2}$. Such coordinate shifts do not change renormalizability or the associated $\b$-functions.}
\be \la{45}
ds^2 = \frac{k}{p^2+ q^2 -1 }( { dp^2 + dq^2 }) = \kk(d\a^2 + \coth^2 \alpha\, d \b^2) \ , \qquad
(p,q) = \cosh\a (\cos \b, \sin \b)
\ . \ee
Another limit is found by
redefining the coordinates and then taking $\kk \to \infty$, $\ka\to 0 \ (b \to \infty)$ as
\be \la{455}
p= \ka \barp\ , \qquad q= 1 + \ka^2 \barq \ , \qquad
\kk \to \infty\ , \quad \ka\to 0 \ , \qquad h\equiv k\ka ={\rm fixed} \ . \ee
In this limit the \lam metric
becomes that of the NAD of $H^2$:
\be \la{46}
ds^2 ={ h\ov \barp ^2 + 2\barq } ( d \barp ^2 + d\barq ^2 ) \ . \ee
Starting from the NAD of $H^2$ \rf{46} we may take the following additional scaling limit
\begin{equation}\la{addsca}
x \to \gamma x \ , \qquad y \to \tfrac12 \sigma \gamma^2 + \gamma y \ , \qquad \gamma \to \infty \ ,
\end{equation}
which gives the abelian T-dual of the $H^2$ metric
\be\la{h2dual}
ds^2 = \frac{h}{x^2 + \sigma}(dx^2 + dy^2) \ .
\ee
For $\sigma = \pm 1$ this gives two different abelian T-duals of the $H^2$ metric
\unskip\foot{The T-dual metric \rf{tdual2} can also be found by taking the limit $\kk\to \infty$, $\ka \to 0$ with
$p = \cosh \chi$, $q = \ka y$ and $h\equiv k\ka$ fixed in the $\l$-model metric \rf{43}.}
\be \la{tdual1}
\sigma = 1 \ , \quad x = \sinh \chi \ , \qquad & ds^2 = h ( d \chi^2 +\frac{1}{\cosh^2\chi} d y^2) \ ,
\\ \la{tdual2}
\sigma = -1 \ , \quad x = \cosh \chi \ , \qquad & ds^2 = h ( d \chi^2 +\frac{1}{\sinh^2\chi} d y^2) \ .
\ee
For $\sigma  = 0$ we find the $H^2$ metric in Poincar\'e patch, which is self T-dual.

The \lam is known to be 1-loop renormalizable with only $\ka$ running \cite{Itsios:2014lca,Appadu:2015nfa}:
\begin{equation}
{\dt }\kk = 0+ \mathcal{O}(\kk^{-1}) \ , \qquad \qquad
\dt{\ka} = - \kk^{-1}(1-\ka^2) + \mathcal{O}(\kk^{-2}) \ .\la{310}
\end{equation}
Our main observation is that to ensure the 2-loop renormalizability of this model one should modify the classical action
\rf{43} by the particular quantum correction like in \rf{7}, which originates from the determinant of integrating over the gauge field.
For the $\frac{SO(1,2)}{SO(2)}$ \lam \rf{42},\rf{43} one finds that the quantum counterterm in \rf{7}
gives the following correction to the classical Lagrangian \rf{43}\foot{In general,
$\det M$ will be depend on the group element $g$ since, written as an operator, $M \propto 1 + b^{-2} P - \Ad_g$.
Let us also note that, somewhat surprisingly,
in this case the 1-loop correction happens to be independent of the deformation parameter $b$.}
\be \la{47}
\Delta L = - \ha ( \del_a \log \det M )^2 \ \to \ - \ha \big[ \del_a \log (p^2 + q^2 -1)\big]^2 \ . \ee
Here $M=M(x)$ is the matrix that appears in the part of the action \rf{40} quadratic in the gauge field, i.e. $L = \ldots + \Tr A_+ M(x)A_-$.

The 1-loop corrected metric of the \lam \rf{43} is therefore given by (cf. \rf{38})
\be \la{48}
ds^2 = \kk\, {\ka\, dp^2 + \ka^{-1}d q^2 \ov p^2+q^2-1} - \frac{1}{2} \Big[\frac{d ( p^2 + q^2-1 )}{p^2 + q^2 -1}\Big]^2
+ \OO ( k^{-1}) \ .
\ee
The 1-loop term preserves the $\mathbb{Z}_2$ symmetry
\begin{equation}
p\leftrightarrow q \ , \qquad \qquad \kappa \to \kappa^{-1} \ ,\la{313}
\end{equation}
of the classical metric \rf{48}.
As a consequence, $\kappa \to \kappa^{-1}$ will be a symmetry of the RG equations.
One can then check directly that the metric \rf{48} solves the 2-loop RG equations \rf{2},\rf{27}
with
\be \la{49}
{\dt }\kk = \kk^{-1}(\ka^{-1}-\ka)^2 + \mathcal{O}(\kk^{-2}) \ , \qquad \qquad
\dt{\ka} = - \kk^{-1}(1-\ka^2) + \mathcal{O}(\kk^{-3}) \ ,
\ee
and the components of the
diffeomorphism vector $\xi_m$ given by
\begin{align}
\xi_p &= -\frac{2p}{p^2+q^2-1}\big[1 + {1 \ov \kk} \frac{\kappa (p^2-1) -\kappa^{-1} p^2 }{p^2+q^2-1} \big] \ ,
\no \\
\xi_q &= -\frac{2q}{p^2+q^2-1}\big[1 + {1 \ov \kk} \frac{\kappa^{-1} (q^2-1) - \kappa q^2)}{p^2+q^2-1} \big] \ . \la{50}
\end{align}
The $\b$-function for the deformation parameter $\ka$
does not receive a 2-loop correction.
Surprisingly, the parameter $k$ (which has the interpretation of the level in the gWZW model) starts running
at 2-loop order.
Still, the RG flow is effectively 1-coupling one as there is an RG invariant (cf. \rf{2888})
\be \la{51}
\dt \rk =0 \ ,\ \ \ \ \qquad \rk \equiv \kk - ({\kappa}^{-1} + {\kappa}) + \mathcal{O}(\kk^{-1}) \ .
\ee
This suggests that it is $\rk$ rather than $k$ that should be identified with level of the gWZW model in \rf{40}.
This corresponds to setting $\kk= \rk+ ({\kappa}^{-1} + {\kappa}) $ in \rf{48}, i.e.
re-interpreting the classical term in \rf{48} multiplied by
${1\ov k} ({\kappa}^{-1} + {\kappa}) $ as an extra quantum counterterm.

In general, the \lam action \rf{40} also contains a WZ term with coefficient $k_{\rm wz}$,
which is not renormalized (and is integer for compact group $G$).
However, for general coefficient $b$ of the deformation term in \rf{40},
the coefficient $k$ of the gWZW part of the action
may not be equal to $k_{\rm wz}$ beyond the classical
level if there is no gauge symmetry relating the two coefficients
as for the gWZW model. In particular, in the case of the 2d target space discussed above, for which the WZ term is trivial,
the shift of $k$ we found in \rf{51} does not appear to contradict any general principle.

These results are consistent with the limits of the \lam discussed above.
Setting (cf. \rf{42},\rf{45})
\be \la{317}
(p,q) = (1 - {\kk}^{-1}) \cosh\a (\cos \b, \sin \b)
\ee
in \rf{48} and taking the gWZW limit, $\kappa \to 1$, we find
\be \la{52}
ds^2 = \kk(d\alpha^2 + \coth^2 \alpha\, d\beta^2)
- 2 (d\alpha^2 - {\coth^2\alpha\ov \sinh^{2}\alpha}\, d \beta^2) + \mathcal{O}(\kk^{-1}) \ , \ee
which precisely matches the large $\kk$ expansion of the exact gWZW metric (cf. \rf{6}), i.e.
\be \la{522}
ds^2 = (\kk-2)\Big(d\alpha^2 + \frac{\coth^2 \alpha\ d\beta^2}{1-\frac{2}{\kk}\coth^2\alpha}\Big) \ . \ee
Note that for $\ka = 1$ the RG invariant in \rf{51} becomes $\rk =\kk-2$, i.e. the
shifted level for the $SO(1,2)/SO(2)$ gWZW model.\foot{In the $\ka = 1$ limit
the diffeomorphism vector \rf{50} becomes a gradient of the exact dilaton of the gWZW model:
$\xi_n = 2 \del_n \Phi$, $ \Phi = - \log (p^2 + q^2 -1) - { 1 \ov 2 k} { 1 \ov p^2 + q^2 -1} + \OO ( k^{-2}) $ (cf.
footnote \ref{f4}). }

The \etm limit requires relating the parameters and coordinates in \rf{38} and in \rf{48}
as follows
\begin{align}
\la{53}
& \ka = \k \big[1+2h^{-1} (1-\k^2)\big] \ , \qquad \qquad h+1+\k^2= \kk \k \ , \\
& p\to \g p \ ,\qquad \qquad q \to \g \k \big[1+ h^{-1} ({1-\k^2}) \big]\, q \ , \qquad \qquad \g\to\infty \ . \la{54}
\end{align}
In the $\g\to \infty$ limit the \lam metric \rf{48} reduces to the \etm metric \rf{38},
while the RG equations \rf{49} give the corresponding ones for the couplings $\k$ and $h$ in \rf{38}
(which are the analytic continuation \rf{344} of \rf{28},\rf{288}), i.e.
\begin{align} \la{555}
\dt h = - (1-\k^2)&\big[1 - h^{-1} (1+\k^2) + \OO(h^{-2})\big] \ , \\
\dt \k = - h^{-1} \k (1-\k^2)&\big[1 - h^{-1} (1+\k^2) + \OO(h^{-2}) \big] \ , \la{55} \end{align}
with the RG invariant \rf{51} given by (cf. \rf{288})
\be
\rk = \kk - ({\ka}^{-1} + {\ka}) + \OO(\kk^{-1}) = \frac{h}{\k} + \OO(h^{-1}) =\nu +\OO(\nu^{-1}) \ . \la{56}
\ee
This may be viewed as a hint that a quantum deformation parameter
that appears
in the corresponding exact S-matrix should be ${\rm q}= e^{ { i \pi \ov \rk} } $ (cf. \cite{Hoare:2015gda}),
which becomes the ${\rm q}= e^{ { i \pi \ov \kk-2} } $ at the gWZW point $\ka=1$.

In the NAD limit \rf{455}
the quantum-corrected metric \rf{48} becomes\foot{Note that since the coordinate redefinition in \rf{455} depends on $\ka=\ka(t)$, 
the diffeomorphism vector in \rf{1} gets an additional contribution compared to \rf{50}.
Indeed, in the NAD limit the components are given by
$ \xi_x = - { 4x \ov x^2 + 2 y} + {1\ov h} {2 x(1 + 3 x^2 + 4 y)\ov ( x^2 + 2 y)^2} + \OO(h^{-2})$,
$\xi_y = - { 2 ( 1 + 2 y) \ov x^2 + 2 y} + {1\ov h} { 2(1 + 2 x^2 + 2 y)\ov ( x^2 + 2 y)^2} + \OO(h^{-2}) .$
}
\be \la{57}
ds^2 = h\, {dx^2 + d y^2 \ov x^2 + 2 y} - \frac{1}{2} \Big[\frac{d ( x^2 + 2 y )}{x^2 + 2 y}\Big]^2 + \OO(h^{-1}) \ , \qquad \qquad h\equiv \kk \ka \ ,
\ee
with the 2-loop $\b$-function for $h$ following from \rf{49},
\be \la{58}
\dt h = - 1 + h^{-1} + \mathcal{O}(h^{-2}) \ , \ee
being the same as in the dual undeformed $H^2$ $\s$-model.

This extends the previous conclusions \cite{Fridling:1983ha,Fradkin:1984ai} about the 1-loop quantum
equivalence of the models related by the non-abelian duality to the 2-loop level.
In particular, \eqref{57} implies that the preservation of quantum equivalence requires a non-trivial 1-loop correction to the
classical NAD model metric.
The origin of this correction can be traced to the finite local
contribution of the determinant \rf{7},\rf{47} that appears when integrating over the 2d gauge field $A_a$ in the path integral
NAD transformation. This counterterm is required to preserve the one-coupling renormalizability of the NAD model at the 2-loop level, or, equivalently, since NAD preserves the classical integrability \cite{Sfetsos:2013wia},
to maintain integrability at the quantum level.

In this way, we have identified how earlier problems checking NAD at 2-loop level \cite{n1,n2,n3,Bonneau:2001za} should be resolved in general.
In particular, as for abelian T-duality, NAD (properly modified by
quantum $\a'$ corrections) should also be a symmetry of the \sm $\beta$-functions or the string effective action to all orders in $\a'$.

Furthermore, the \etm and the \lam are, in general, related \cite{Vicedo:2015pna,Hoare:2015gda,Sfetsos:2015nya,Klimcik:2015gba,Hoare:2017ukq}
by the Poisson-Lie (PL) duality \cite{Klimcik:1995dy} (and analytic continuation).
Therefore, the same observations made above for NAD should apply also to PL duality,
which (with suitable quantum corrections) should be a symmetry not only at 1-loop order \cite{Valent:2009nv,Sfetsos:2009dj}, but also at higher loops.

Finally, let us observe that taking the following 1-loop modification of the limit \rf{addsca} (cf. \rf{317})
\be\la{329}
x = \gamma (1-h^{-1}) x \ , \qquad y \to \tfrac12 \sigma \gamma^2 + \gamma (1-h^{-1}) y \ , 
\qquad \gamma \to \infty \ ,
\ee
in the quantum-corrected NAD metric \rf{57}, with an additional shift of the coupling $h \to h +2$, leads to
\be\la{327}
ds^2 = \frac{h}{x^2 + \sigma} \Big[dx^2 + \big(1+ h^{-1}  \frac{2x^2}{x^2+\sigma}\big) dy^2\Big] + \OO(h^{-1}) \ ,
\ee
or explicitly for $\sigma=-1$ (cf. \rf{tdual2}) $ds^2 = h \big[d\chi^2 +\frac{1}{\sinh^2 \chi} \big(1+2 h^{-1}\coth^2\chi\big)dy^2\big] + \OO(h^{-1})$. 
Here the  1-loop correction to the abelian T-dual of the $H^2$ metric \rf{h2dual}  matches that which follows
from the known 2-loop modification of the T-duality rule \rf{5} \cite{Tseytlin:1991wr}:
since the $H^2$ metric dual to \rf{h2dual} is not deformed, here ${\cal M} = x^2 + \sigma$ and $\tilde {\cal M}$ corresponds to \rf{327}.
Note that this effectively explains 
the origin of this modification \rf{5} in this special case where the T-dual metric
can be found as a limit 
of a NAD model:
  it comes from a combination of the  quantum correction  in \rf{57}, an order $h^{-1}$ 
 coordinate transformation  in \rf{329}, and additional quantum counterterm  proportional to the classical metric 
 that comes from    the shift $h \to h +2 $.

\section{Concluding remarks}

In this paper we have demonstrated that, just as for gauged WZW models,
the invariance of integrable $\s$-models under the
two-loop (and higher) RG flow requires a specific quantum
deformation of the classical Lagrangian (i.e. of the target space metric and $B$-field).

In particular, we proposed an exact metric for the $\eta$-deformation of the $S^2$ (or $H^2$) model
that solves the 3-loop RG flow equations and is consistent with the gWZW limit.
We also found the leading-order deformation of the \lam
for $SO(1,2)\ov SO(2)$, which solves the
2-loop RG equations and is consistent with the gWZW and $\eta$-model limits.
For the $\l$-model we identified the origin of the deformation as a finite counterterm
resulting from the determinant of integrating over the auxiliary 2d gauge field
(the same determinant that leads to the correction of the dilaton term on a curved 2d background).
As a by-product, this implies a
resolution of the earlier problem in checking the consistency of non-abelian duality at the 2-loop level.
Similar observations should apply
to generic YB deformations of \sms and to PL duality.

Among various open problems let us mention the construction of an
exact generalization of the metric \rf{48} for the \lam that reduces in the appropriate limits to the exact gWZW
metric \rf{522} and the \etm metric \rf{36},\rf{37}. This is non-trivial given the lack of isometries in the \lam metric
and that the effective action approach used for the gWZW model \cite{Tseytlin:1993my} does not appear to apply directly to the non-conformal 
$\l$-model \rf{40}.

It would also be important to repeat a similar analysis for \sms with 3-dimensional target spaces,
i.e. for integrable deformations of $S^3$ or $H^3$. In addition, one may consider the non-abelian dual of $S^3$
and confirm that the modifications of $G_{mn}$ and $B_{mn}$ required to solve the 2-loop RG equations
again originate from a determinant such as \rf{7}.

It would be interesting to see if the exact metric \rf{24} of the $\eta$-model,
translated to the Hamiltonian framework, implies some natural interpretation in terms of a deformation of
the classical integrable structure.
Finally, it would also be interesting to establish a
connection of the quantum-deformed ``sausage" \sm to the massive non-perturbative S-matrix
\cite{Fateev:1992tk} and possibly to the dual massive 2d QFT
(cf.  \cite{fateev96,Litvinov:2018bou}).

\section*{Acknowledgments}

We would like to thank R. Borsato, J.L. Miramontes, F. Seibold, K. Sfetsos and L. Wulff for
useful discussions.
BH was supported by grant no. 615203 from the European Research Council under the FP7
and by the Swiss National Science Foundation through the NCCR SwissMAP.
NL was supported by the EPSRC grant EP/N509486/1.
AAT was supported by the STFC grant ST/P000762/1.

\bigskip

\appendix

\section{3-loop \texorpdfstring{$\b$}{beta}-function in different renormalization schemes} \label{A}
\def\theequation{A.\arabic{equation}}
\setcounter{equation}{0}

In general, in quantum field theories
changes of a renormalization scheme are equivalent to local redefinitions of the coupling constants,\foot{Here we reinstate the loop-counting parameter $\a'=\hbar$.}
\be
g^i \to \hat{g}^i (g ) = g^i+ \a' \, c^i_{jk} g^j g^k + \ldots \ . \la{redef}
\ee
Under \rf{redef} the $\b$-functions $\b^i \equiv \dt g^i$ transform as a contravariant vector in the space of couplings:
\be \la{deltab}
\delta \b^i \equiv\hat{\beta}^i - \beta^i = \frac{\del \, \delta g^i}{\del g^j} \b^j - \frac{\del \b^i }{\del g^j} \delta g^j + \ldots \ , \qquad \qquad \delta g^i \equiv \hat{g}^i - g^i \ ,
\ee
where the higher order corrections are non-linear in $\delta g$. Assuming $\b^i$ and $\delta g^i$ both start at 1-loop order, i.e.
are both $\O(\a')$, then $\delta \b$ is $\O(\a'^2)$, so the 1-loop $\b$-function is scheme-invariant.
In single-coupling theories, one can further show that the 2-loop $\b$-function is also scheme-invariant.

Now let us turn to the case of 2d \sms for which the coupling redefinitions \rf{redef} are local redefinitions of the target space geometry.
In order to preserve the manifestly covariant structure of the action \rf{1} we restrict to covariant redefinitions.
Restricting further to include only redefinitions with a natural interpretation in terms of changing the subtraction scheme,\foot{While the term $\a'^2 \nabla^2 R_{mn}$ in \rf{metricredef} has no natural interpretation as a change of subtraction scheme starting from the minimal subtraction scheme, it is required in order that the transformation
\rf{2dredef} be closed under inversion in the case of the \sm with a 2d target space,
i.e. in order to be able to move back to the minimal subtraction scheme starting from another scheme.} we consider $G_{mn} \to
\hat{G}_{mn}$ with
\begin{align} \la{metricredef}
&
\hat{G}_{mn} = G_{mn} + c \, \a' R_{mn} + d \, \a'^2 (R^2)_{mn} + e \, \a'^2 \nabla^2 R_{mn} + \O(\a'^3) \ , \\
&(R^2)_{mn} \equiv R_{mabc} {R_n}^{abc} \ , \no
\end{align}
where $c, d, e$ are arbitrary coefficients (for simplicity we consider the special case with vanishing $B$-field; for a general discussion see \cite{mt}).

At the leading $\O(\a')$ order, both $\beta_{mn} = \a' R_{mn} + \ldots$ and $\delta G_{mn} = c\, \a' R_{mn}+\ldots $ only contain a single term proportional to $R_{mn}$.
Thus at the leading $\O(\a'^2)$ order in \rf{deltab}, their contributions cancel and the 2-loop $\b$-function is invariant.\foot{Note that if one also included redefinitions of the form $\a' R G_{mn}$ then this argument would not go through in general dimensions. However, for the special case of a 2d target space, the identity $R_{mn} = \ha R G_{mn}$ means these terms are proportional and the 2-loop $\b$-function remains invariant.}

Specializing further to the case of a 2d target space, the redefinition \rf{metricredef} becomes
\begin{align} \la{2dredef}
&
\hat{G}_{mn} = G_{mn} \Big[ 1 + \ha c \, \a' R + \ha d \, \a'^2 R^2 + \ha e \, \a'^2 \nabla^2 R \Big] \ .
\end{align}
Possible scheme dependence starts at 3-loop order, where the relevant redefinitions are those written in \rf{metricredef}.
The most general form of a covariant expression for a 3-loop $\b$-function
is\foot{In \rf{3loopbeta} we have dropped terms of the form $\nabla_{(m} V_{n)}$ since these may be absorbed into the diffeomorphism vector term in \rf{2}. We have excluded the term $\a'^2 \nabla^2 \nabla^2 R G_{mn}$ in \rf{3loopbeta} since it does not arise naturally through changes of subtraction scheme starting from the minimal subtraction scheme.}
\be \la{3loopbeta}
\b_{mn} = \Big[ \ha \a' R + \tfrac{1}{4} \a'^2 R^2 + c_1 \, \a'^3 R^3 + c_2 \, \a'^3 (\nabla R)^2 + c_3 \, \a'^3 R \nabla^2 R \Big] G_{mn}
+ c_4 \, \a'^3 \nabla_m R \nabla_n R \ . \ee
Not all of the coefficients $c_1,\ldots ,c_4$ can be scheme-dependent
as legitimate schemes should be parametrized by the 3-dimensional space of coupling redefinitions \rf{metricredef}.
To find the values of $c_i$ that are allowed to appear in the $\b$-function in a particular scheme
let us start from the expression for the $\b$-function \rf{3loopbeta} in the
minimal subtraction scheme where \cite{Graham:1987ep,Foakes:1987ij,Foakes:1987gg}
\be
\qquad c_1 = \tfrac{5}{32} \ , \qquad c_2 = \tfrac{1}{16} \ , \qquad c_3 = 0\ , \qquad c_4=-\tfrac{1}{16}\ .
\ee
Implementing the metric redefinition \rf{metricredef} we find from \rf{deltab} that the coefficients in the
transformed $\b$-function are given by
\begin{align}
&\hat{c}_1 = \tfrac{5}{32} +\tfrac{1}{8}(c - 2d) \ , && \hat{c}_2 = \tfrac{1}{16} -\tfrac{1}{4}(c - 2d)-\tfrac{1}{8}(4e+c^2)\ , \no \\
& \hat{c}_3 = -\tfrac{1}{8}(4e+c^2)\ , && \hat{c}_4=-\tfrac{1}{16}\ . \la{genscheme}
\end{align}
We note that since \rf{genscheme} only depends on the redefinition parameters $c,d,e$ through the combinations $c-2d$ and $4e+c^2$, the choice of $c,d,e$ to give any particular values of $c_1,\ldots,c_4$ is not unique, but rather there is one free parameter. In the case of the \etm metric studied in section 2, the ``natural" scheme \rf{31} is given by \rf{genscheme} with, e.g., $(c,d,e)=(0, \tfrac{5}{8}, \ha)$; i.e. it is related to the minimal subtraction scheme by the redefinition
\begin{align}
G_{mn}^{(\rm nat)}= \big[ G_{mn} + \tfrac{5}{8} \, \a'^2 ({R}^2)_{mn} + \ha \, \a'^2 \nabla^2 R_{mn} \big] ^{(\rm min)}\ .
\end{align}



\end{document}